# 28nm Fully-Depleted SOI Technology: Cryogenic Control Electronics for Quantum Computing

H. Bohuslavskyi[1,2], S. Barraud[1], M. Cassé[1], V. Barral[1], B. Bertrand[1], L. Hutin[1],
F. Arnaud[3], P. Galy[3], M. Sanquer[2], S. De Franceschi[2], and M. Vinet[1]

[1] CEA, LETI, Minatec Campus, F-38054 Grenoble, France ; [2] CEA, INAC-PHELIQS, F-38054 Grenoble, France ; [3] STMicroelectronics, 850 rue J. Monnet, 38920 Crolles, France.
E-mail: sylvain.barraud@cea.fr

## Abstract

This paper reports the first cryogenic characterization of 28nm Fully-Depleted-SOI CMOS technology. A comprehensive study of digital/analog performances and body-biasing from room to the liquid helium temperature is presented. Despite a cryogenic operation, effectiveness of body-biasing remains unchanged and provides an excellent $V_{TH}$ controllability. Low-temperature operation enables higher drive current and a largely reduced subthreshold swing (down to 7mV/dec). FDSOI can provide a valuable approach to cryogenic low-power electronics. Applications such as classical control hardware for quantum processors are envisioned.

## Introduction

Since the early 2010's, increasing attention has been paid to Si-based spin qubits for quantum computing [1-3]. Besides allowing for long spin coherence, largely enhanced by the use of isotopically enriched nuclear spin-free $^{28}$Si layers [3], Si qubits benefit from well-established microelectronics fabrication techniques. This constitutes a clear asset in the prospect of large-scale qubit integration [4]. Recently, the implementation of Si spin qubit on a foundry-compatible CMOS SOI platform was demonstrated [5, 6], marking an important first step towards the realization of a Si-based quantum computer. In a quantum processor, qubits need to be individually addressed, i.e. initialized, manipulated, and measured. This large-scale parallelism seems hardly manageable without the use of co-integrated, or at least proximal, classical hardware. Therefore, extending the operation of both digital and analog Si electronics to temperatures as low as 4 K (and below) appears as an urgent task to undertake in parallel with qubit development. At such low temperatures, however, detrimental effects on device operation have been observed due to dopant freeze-out and charge trapping [7-8]. Nevertheless, the actual impact of these effects in advanced CMOS technology remains only barely explored [9]. Here, for the first time, we present a study of digital/analog performances and body-biasing of FDSOI down to 4K. We aim at assessing the potential of FDSOI for low-temperature electronics. FDSOI is an attractive solution since it can simultaneously deliver low-power electronics and a versatile technology framework for qubits development [4-6].

# FDSOI process features

The FDSOI transistors are fabricated with a gate-first high-κ metal gate. They are processed on 300mm (100) SOI wafers with a buried oxide thickness of 25nm **[10]**. The 7nm-thick Si channel is undoped. The EOT is 1.55nm for NMOS (resp. 1.7nm for PMOS) for thin oxide (GO1) and 3.7nm for thick oxide (GO2) transistors. Regular $V_{TH}$ transistors are studied with PWELL (resp. NWELL) and additional P-type (resp. N-type) ground plane implantation for NMOS (resp. PMOS) devices. **Fig.1** summarizes the main technology features and shows a cross-section of FDSOI transistor.

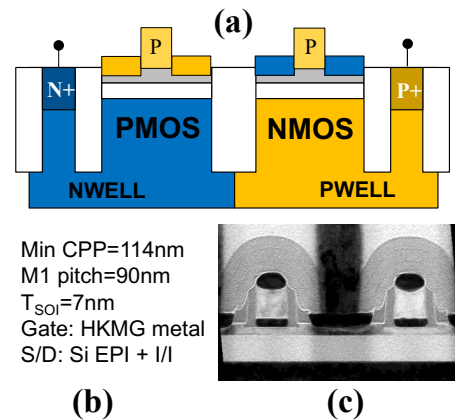

Fig.1. (a) Cross-section, (b) key ground rules, technology features and (c) TEM of FDSOI transistor.

# Device digital performances down to 4K

The carrier mobility ($\mu_{EFF}$) *vs* inversion charge ($N_{INV}$) is an important parameter to characterize the behavior of devices. $N_{INV}$ is obtained by integration of gate-channel ($C_G$) capacitance shown in **Fig.2** for different temperature (T). The T-dependence results in a shift in $V_{TH}$ but does not influence much the shape of $C_G$. The electron and hole $\mu_{EFF}$ measured down to 4K are shown in **Fig. 3**.

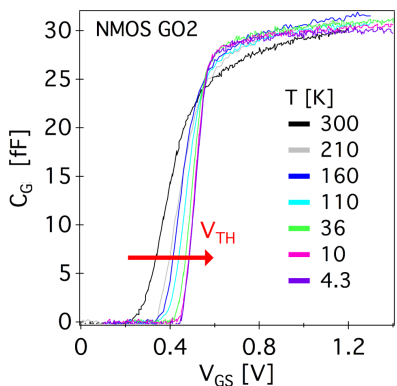
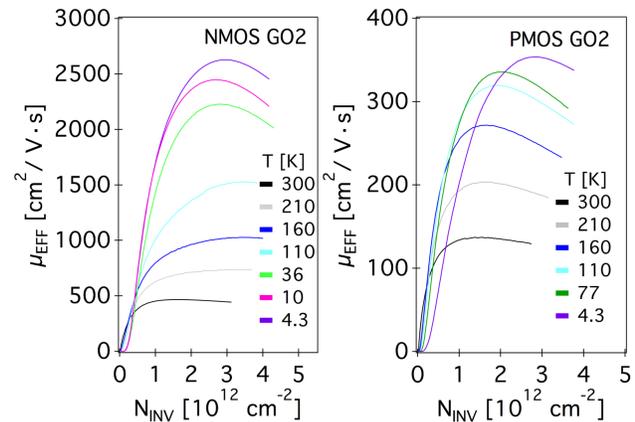

Fig.2. Gate-channel capacitance versus $V_{GS}$ for different T ($L_G$=W=2μm, $V_{BACK}$=0V).

Fig.3. Electron and hole $\mu_{EFF}$ versus $N_{INV}$ for 4K≤T≤300K ($L_G$=W=2μm, $V_{BACK}$=0V).

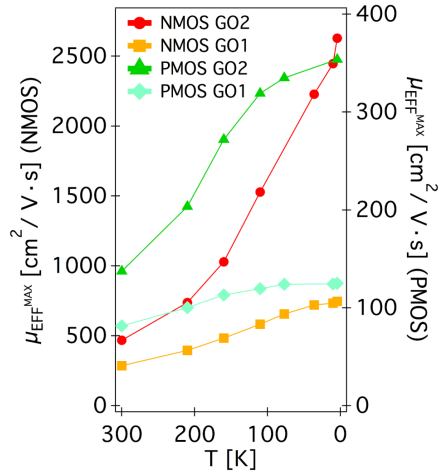

Fig.4. $\mu_{EFF,MAX}$ versus T. $L_G=W=2\mu m$ for GO2 and $L_G=W=1\mu m$ for GO1. $V_{BACK}=0V$

Since phonon mobility is sufficiently weak and can be neglected at such low temperature, the carrier mobility is enhanced as temperature decreases (**Fig.4**). It can also be noted that thick gate oxide (GO2) weakens remote Coulomb and soft-optical phonon scattering. It therefore leads to an improved carrier mobility **[11,12]**. Typical transfer characteristics of GO2 transistors are plotted in **Fig.5** (NMOS) and **Fig.6** (PMOS). The increase of $|V_{TH}|$ as the temperature is reduced is accompanied by a significant improvement of current for a given overdrive gate voltage ($|V_{GS}-V_{TH}|=0.5V$). At 4K, a subthreshold swing $SS_{LIN}$ of only 7mV/dev is achieved in the linear regime (**Fig.7**). Even at high drain voltage ($V_{DS}=0.9V$), the evolution of subthreshold swing versus T remains the same (see inset of **Fig.7**). At the same time, the linear current $I_{ODLIN}$ of NMOS GO2 is enhanced by a factor ×6.2 (**Fig.8**). By comparison with 300K, the substantial increase of devices performance at low-temperature is summarized in **Fig.9**. The saturation current $I_{ODSAT}$ ($|V_{GS}-V_{TH}|=0.5V$ and $V_{DS}=0.9V$) increases due to higher mobility while the subthreshold swing $SS_{SAT}$ remains very low. GO1 transistors (N/PMOS) show a higher saturation current than GO2 transistors due to a lower EOT.

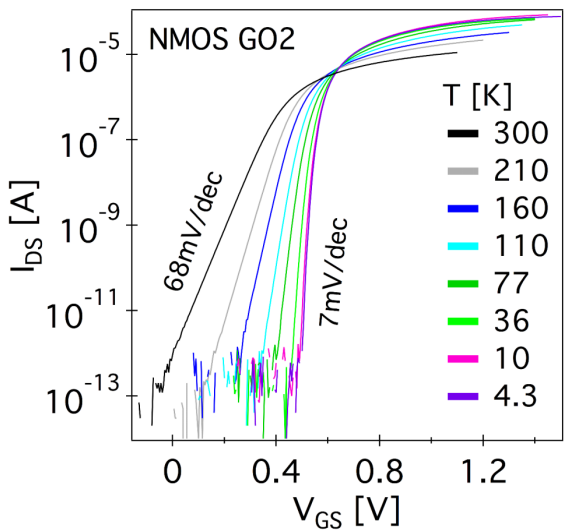

Fig.5. $I_{DS}(V_{GS})$ for different T ($L_G=W=2\mu m$, $V_{DS}=50mV$, $V_{BACK}=0V$).

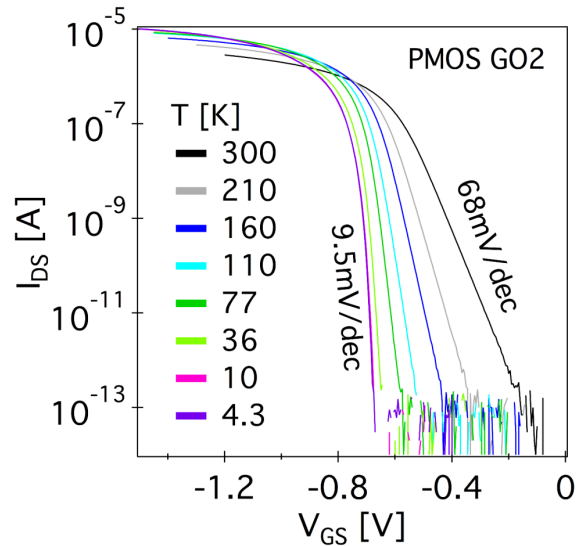

Fig.6. $I_{DS}(V_{GS})$ for different T ($L_G=W=2\mu m$, $V_{DS}=50mV$, $V_{BACK}=0V$).

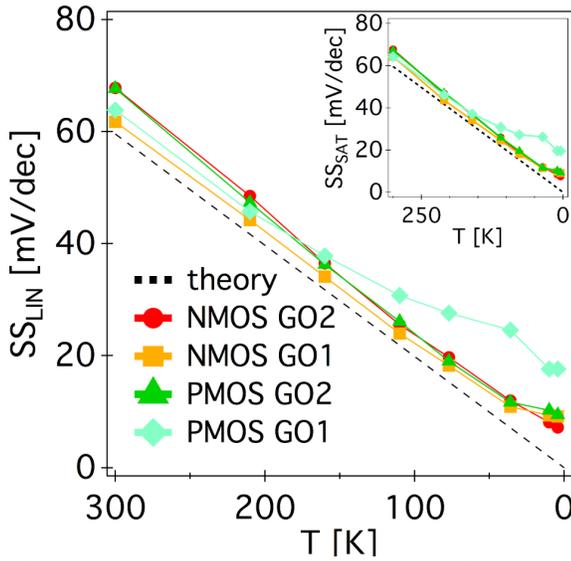

Fig.7. $SS_{LIN}$ versus T ($L_G$=W=2µm for GO2 and $L_G$=W=1µm for GO1).

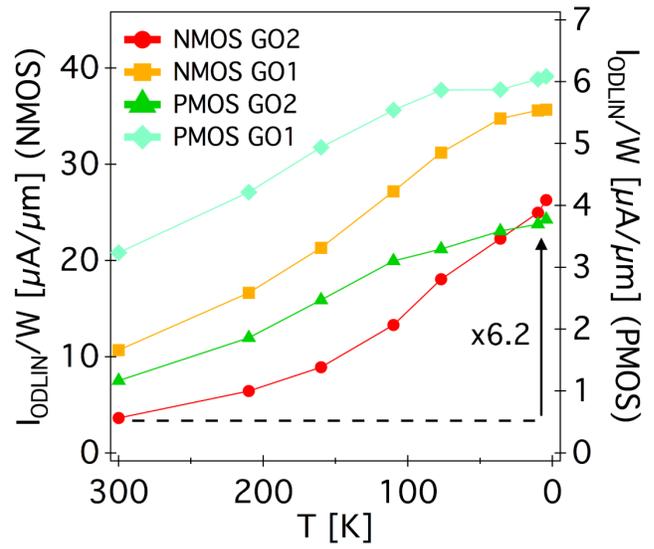

Fig.8. $I_{ODLIN}$ versus T ($L_G$=W=2µm for GO2 and $L_G$=W=1µm for GO1). $V_{BACK}$=0V.

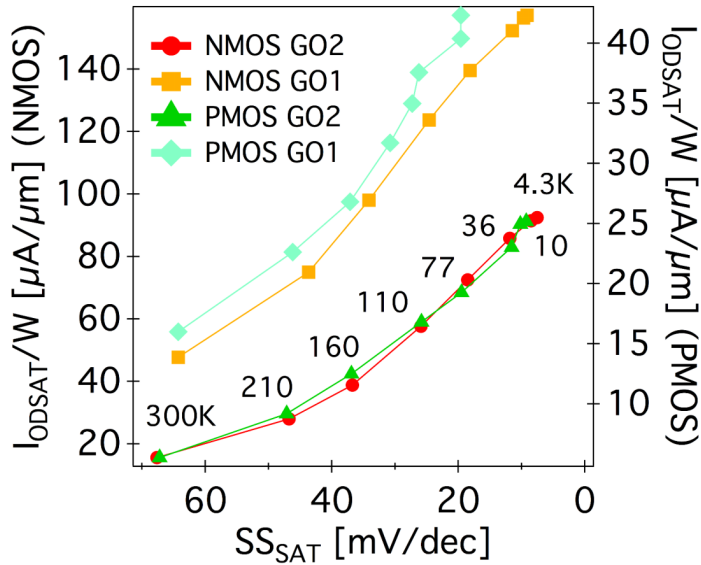

Fig.9. $I_{ODSAT}$ versus $SS_{SAT}$ for different T ($L_G$=W=2µm for GO2 and $L_G$=W=1µm for GO1). $V_{BACK}$=0V.

## Back bias efficiency down to 4K

The use of back biasing (BB) is a key factor in optimizing performances and power consumption. Therefore, maintaining its efficiency at very low temperatures is crucial. In addition, back-biasing can provide a useful control tool for qubits, enabling fast readout circuitry and a tunable coupling between neighboring quantum dots [4,6]. The $V_{TH}$ controllability offered by the back-gate at 4K is shown in **Fig.10**. Compared to 300K, the BB factor ($\gamma=\Delta V_{TH}/\Delta V_{BACK}$) for N and PMOS remains unchanged (**Fig.11**). It means that the device behavior is not altered by freeze-out phenomena who are likely to play an important role at low-temperature [8,9].

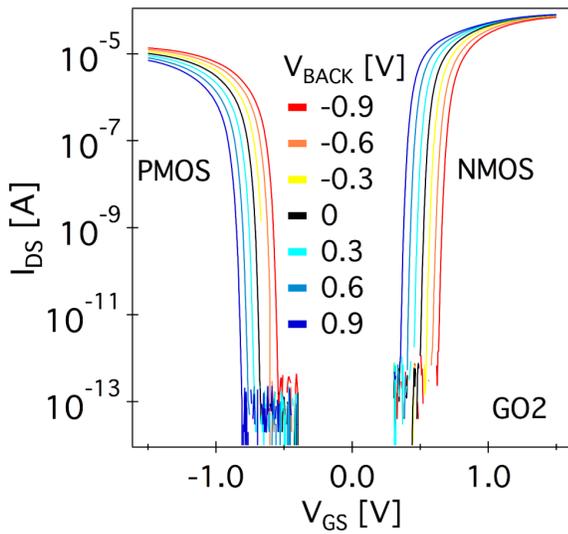
Fig.10. $I_{DS}(V_{GS})$ for different $V_{BACK}$ ($L_G=W=2\mu m$, $V_{DS}=50mV$, $T=4.3K$).

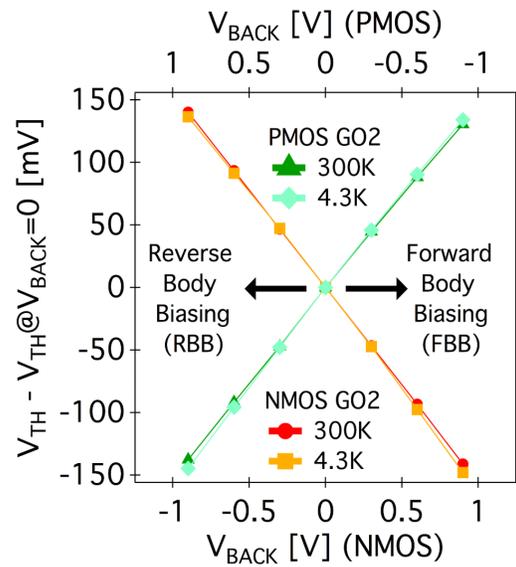
Fig.11. $V_{TH,NORM}$ versus $V_{BACK}$ (GO1/GO2 devices, $V_{DS}=50mV$).

At short gate length ($L_G=28nm$), the device performance (**Fig.12**) studied on NMOS GO1 transistors leads to a similar conclusion. The subthreshold swing is strongly reduced with T independently of the active layer width W (**Fig.13**). **Fig14** shows that at 28nm gate length, the linear current ($|V_{GS}-V_{TH}|=0.5V$) is significantly improved (+62% at 4K). The effect of BB measured at 4K is identical to the 300K one (**Fig.15**). In case of PMOS GO1, low-temperature combined with a short $L_G$ lead to a modification of subthreshold behavior: oscillations occur at low $V_{DS}$ (**Fig.16**). They are most likely induced by the presence of dopants diffused from S/D into the channel **[13]**. However, these oscillations are strongly suppressed at high $V_{DS}$ (**Fig.17**).

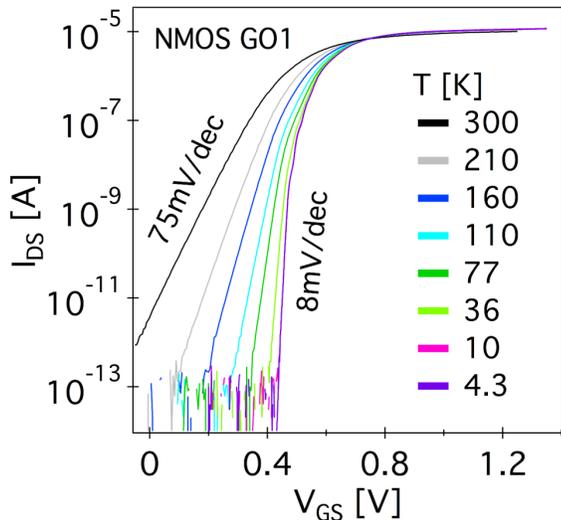
Fig.12. $I_{DS}(V_{GS})$ for different T ($V_{DS}=50mV$, $V_{BACK}=0V$, $L_G=28nm$, $W=80nm$).

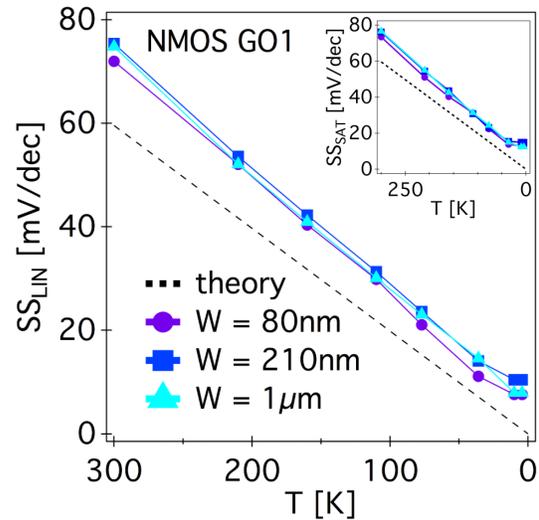
Fig.13. $SS_{LIN}$ versus T for NMOS with various W ($L_G=28nm$, $V_{DS}=50mV$, $V_{BACK}=0V$). $SS_{SAT}(T)$ is in inset.

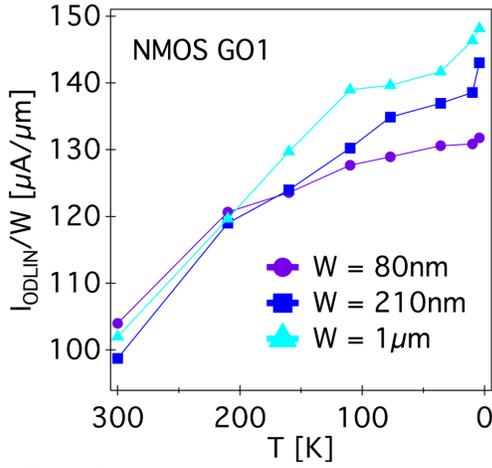

Fig.14. $I_{ODLIN}$ versus T for different W ($L_G$=28nm, $V_{DS}$=50mV, $V_{BACK}$=0V).

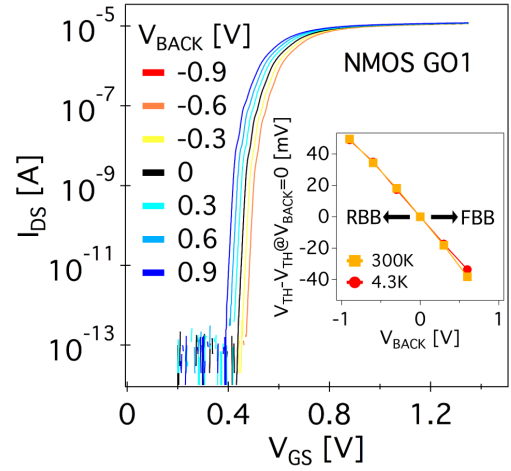

Fig.15. $I_{DS}(V_{GS})$ for different $V_{BACK}$ ($L_G$=28nm, $V_{DS}$=50mV, T=4K). The effect of BB is shown in the inset.

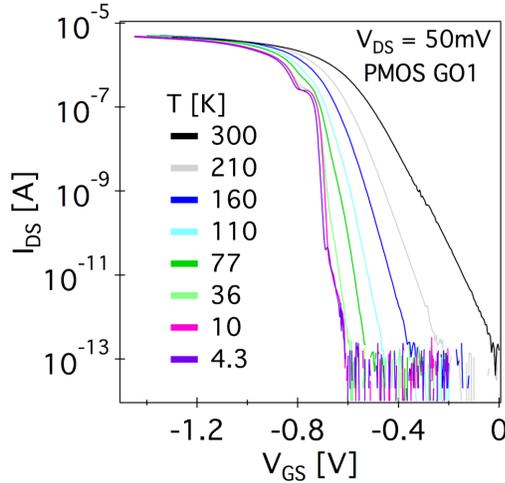

Fig.16. $I_{DS}(V_{GS})$ for different T ($V_{BACK}$=0V, $L_G$=28nm and W=80nm).

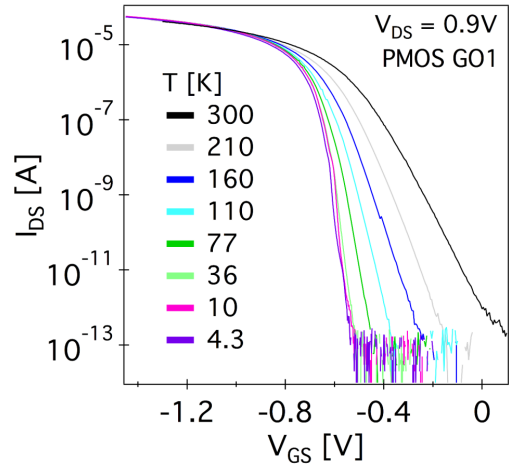

Fig.17. $I_{DS}(V_{GS})$ for different T ($V_{BACK}$=0V, $L_G$=28nm and W=80nm).

## Device analog performance down to 4K

Device analog performance is also assessed down to 4K through the key parameters such as the transconductance $G_M$, the output conductance $G_D$ and the intrinsic gain $A_{V0}$= $G_M/G_D$. The figure of merit $G_M/I_{DS}$ vs $I_{DS} \times L/W$ for NMOS (resp. PMOS) GO1 is reported in **Fig.18** (resp. **Fig.19**). $G_M/I_{DS}$ being inversely proportional to the subthreshold swing (in weak inversion) and proportional to $\mu_{EFF}$ (in strong inversion) the [$G_M/I_{DS}$ vs $I_{DS}$] metric at 4K is strongly improved. $G_M/W$ vs intrinsic voltage gain ($A_{V0}$) is shown in **Fig.20**. Low-temperature and back-biasing significantly improve the transconductance $G_M$ (due to increased $\mu_{EFF}$) and the intrinsic gain $A_{V0}$ (due to reduced $G_D$).

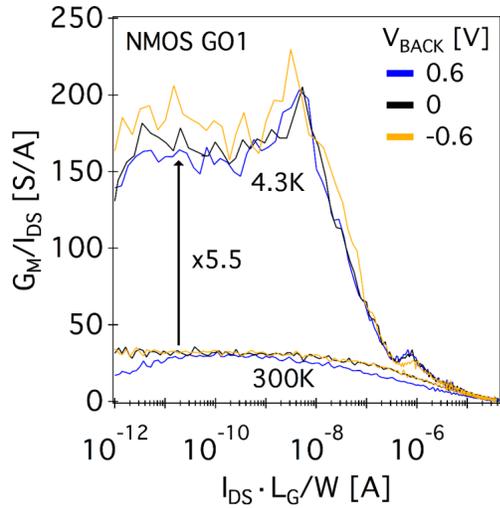

Fig.18. $G_M/I_{DS}$ vs $I_{DS} \times L_G/W$ for different $V_{BACK}$, T. NMOS GO1 device with $\underline{L_G=28nm}$ and W=80nm.

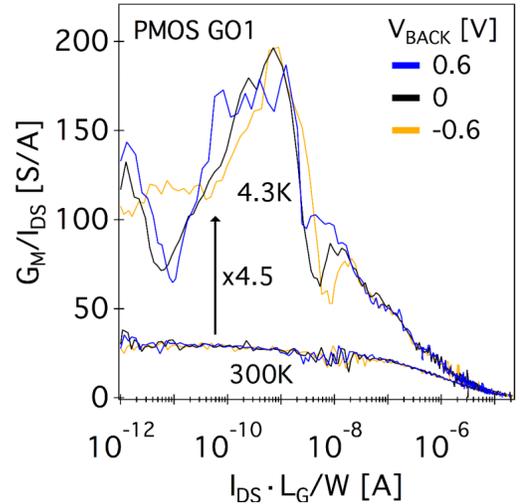

Fig.19. $G_M/I_{DS}$ vs $I_{DS} \times L_G/W$ for different $V_{BACK}$, T. PMOS GO1 device with $\underline{L_G=28nm}$ and W=80nm.

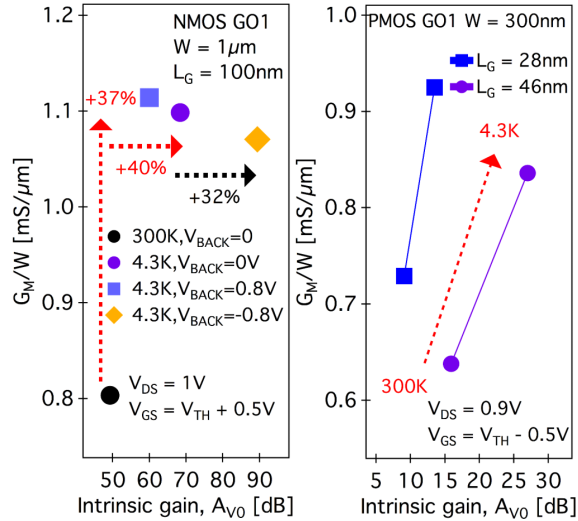

Fig.20. $G_M/W$ vs $A_{V0}$ for different T, W, $L_G$ and $V_{BACK}$. N/PMOS GO1 device.

## Conclusions

For the first time, the operation of 28nm Fully-Depleted-SOI CMOS technology was successfully demonstrated down to 4K. We have shown that a reduced temperature operation largely improves CMOS device performance. Additionally, under a DC operation, the ability to adjust $V_{TH}$ through back-biasing remains unchanged at 4K. This unique capability could be particularly helpful in the development of fast power-efficient peripheral circuitry (multiplexing, control and readout of qubits) and allows increased flexibility in the design of a scalable classical-quantum interface.

## Acknowledgements

This work is partly funded by French Public Authorities (NANO 2017) and Equipex FDSOI. We also acknowledge financial support from the EU under Project MOS-QUITO (No.688539).

## References

**[1]** E. Kawakami *et al.*, *Nature Nanotech.*, 9, p.666, 2014. **[2]** J.T. Muhonen *et al., Nature Nanotech.*, 9, p.986, 2014. **[3]** M. Veldhorst *et al.*, *Nature Nanotech.*, 9, p.981, 2014. **[4]** S. De Franceschi *et al.*, *EUROSOI-ULIS*, p.124, 2016, **[5]** L. Hutin *et al.*, p.1-2, *VLSI*, 2016. **[6]** De Franceschi *et al.*, *IEDM,* 2016. **[7]** G. Ghibaudo *et al*., *Microelectronics Eng*., 19, p.833, 1992. **[8]** E. Simoen *et al.*, *IEEE TED*, 42, p1100, 1995. **[9]** E. Charbon *et al.*, *IEDM*, 2016. **[10]** N. Planes *et al.*, *VLSI,* p.133, 2012. **[11]** M. Cassé *et al.*, *IEEE TED*, 53, p.759, 2006. **[12]** V.-H. Nguyen *et al*., *IEEE TED*, 61, p.3096, 2014. **[13]** R. Wacquez et al., *VLSI*, p.193, 2010.